\documentclass[12pt]{article}
\usepackage{html}
\usepackage{epsf}
\usepackage{graphicx}
\usepackage{amssymb}
\begin{document}
\title{MATTERS OF GRAVITY, The newsletter of the APS Topical Group on 
Gravitation}
\begin{center}
{ \Large {\bf MATTERS OF GRAVITY}}\\ 
\bigskip
\hrule
\medskip
{The newsletter of the Topical Group on Gravitation of the American Physical 
Society}\\
\medskip
{\bf Number 33 \hfill Winter 2009}
\end{center}
\begin{flushleft}
\tableofcontents
\vfill\eject
\section*{\noindent  Editor\hfill}
David Garfinkle\\
\smallskip
Department of Physics
Oakland University
Rochester, MI 48309\\
Phone: (248) 370-3411\\
Internet: 
\htmladdnormallink{\protect {\tt{garfinkl-at-oakland.edu}}}
{mailto:garfinkl@oakland.edu}\\
WWW: \htmladdnormallink
{\protect {\tt{http://www.oakland.edu/physics/physics\textunderscore people/faculty/Garfinkle.htm}}}
{http://www.oakland.edu/physics/physics_people/faculty/Garfinkle.htm}\\

\section*{\noindent  Associate Editor\hfill}
Greg Comer\\
\smallskip
Department of Physics and Center for Fluids at All Scales,\\
St. Louis University,
St. Louis, MO 63103\\
Phone: (314) 977-8432\\
Internet:
\htmladdnormallink{\protect {\tt{comergl-at-slu.edu}}}
{mailto:comergl@slu.edu}\\
WWW: \htmladdnormallink{\protect {\tt{http://www.slu.edu/colleges/AS/physics/profs/comer.html}}}
{http://www.slu.edu//colleges/AS/physics/profs/comer.html}\\
\bigskip
\hfill ISSN: 1527-3431

\bigskip

DISCLAIMER: The opinions expressed in the articles of this newsletter represent
the views of the authors and are not necessarily the views of APS.

\begin{rawhtml}
<P>
<BR><HR><P>
\end{rawhtml}
\end{flushleft}
\pagebreak
\section*{Editorial}

The next newsletter is due September 1st.  This and all subsequent
issues will be available on the web at
\htmladdnormallink 
{\protect {\tt {http://www.oakland.edu/physics/Gravity.htm}}}
{http://www.oakland.edu/physics/Gravity.htm} 
All issues before number {\bf 28} are available at
\htmladdnormallink {\protect {\tt {http://www.phys.lsu.edu/mog}}}
{http://www.phys.lsu.edu/mog}

Any ideas for topics
that should be covered by the newsletter, should be emailed to me, or 
Greg Comer, or
the relevant correspondent.  Any comments/questions/complaints
about the newsletter should be emailed to me.

A hardcopy of the newsletter is distributed free of charge to the
members of the APS Topical Group on Gravitation upon request (the
default distribution form is via the web) to the secretary of the
Topical Group.  It is considered a lack of etiquette to ask me to mail
you hard copies of the newsletter unless you have exhausted all your
resources to get your copy otherwise.

\hfill David Garfinkle 

\bigbreak

\vspace{-0.8cm}
\parskip=0pt
\section*{Correspondents of Matters of Gravity}
\begin{itemize}
\setlength{\itemsep}{-5pt}
\setlength{\parsep}{0pt}
\item John Friedman and Kip Thorne: Relativistic Astrophysics,
\item Bei-Lok Hu: Quantum Cosmology and Related Topics
\item Gary Horowitz: Interface with Mathematical High Energy Physics and
String Theory
\item Beverly Berger: News from NSF
\item Richard Matzner: Numerical Relativity
\item Abhay Ashtekar and Ted Newman: Mathematical Relativity
\item Bernie Schutz: News From Europe
\item Lee Smolin: Quantum Gravity
\item Cliff Will: Confrontation of Theory with Experiment
\item Peter Bender: Space Experiments
\item Jens Gundlach: Laboratory Experiments
\item Warren Johnson: Resonant Mass Gravitational Wave Detectors
\item David Shoemaker: LIGO Project
\item Stan Whitcomb: Gravitational Wave detection
\item Peter Saulson and Jorge Pullin: former editors, correspondents at large.
\end{itemize}
\section*{Topical Group in Gravitation (GGR) Authorities}
Chair: David Garfinkle; Chair-Elect: 
Stan Whitcomb; Vice-Chair: Steve Detweiler. 
Secretary-Treasurer: Gabriela Gonzalez; Past Chair:  Dieter Brill;
Delegates:
Alessandra Buonanno, Bob Wagoner,
Lee Lindblom, Eric Poisson,
Frans Pretorius, Larry Ford.
\parskip=10pt

\vfill
\eject

\section*{\centerline
{GGR program at the APS meeting in Denver}}
\addtocontents{toc}{\protect\medskip}
\addtocontents{toc}{\bf GGR News:}
\addcontentsline{toc}{subsubsection}{
\it GGR program at the APS meeting in Denver, by David Garfinkle}
\parskip=3pt
\begin{center}
David Garfinkle, Oakland University
\htmladdnormallink{garfinkl-at-oakland.edu}
{mailto:garfinkl@oakland.edu}
\end{center}
We have an exciting GGR related program at the upcoming APS ``April'' meeting
(actually May 2-5) in Denver.  Our chair-elect
Stan Whitcomb did an excellent job of putting together this program.  
At the APS April meeting there will be several invited sessions of talks sponsored by the Topical Group in Gravitation (GGR).  
The large number of sessions sponsored by GGR means that our Topical Group 
has become one of the most important units at this meeting: only the Divisions
of Astrophysics, Particles and Fields, and Nuclear Physics have a larger 
presence at the April meeting.  
In addition, there will be plenary talks on gravitational topics, 
and several of the invited sessions sponsored by other APS units are 
likely to be of interest to GGR members.\\  

2009 is the 400th anniversary of Galileo's telescope, and this meeting emphasizes new ways of looking at the universe. Plenary talks include:\\

First Results from Fermi/GLAST  Peter Michelson (Stanford University)\\
Merging Black Holes  Joan Centrella (NASA/Goddard)\\
Nature's Highest Energy Messenger: Pierre Auger Observatory\\
James Cronin (Univ. of Chicago)\\

The invited sessions sponsored by GGR are as follows:\\

Merging Galaxies\\
(joint with DAP)\\
Stelios Kazantzidis Fueling of Black Holes in Galaxy Mergers\\
Marta Volonteri Cosmological Growth of Black Holes\\
Alberto Vecchio Probing Black Hole Mergers with LISA\\

Numerical simulations of coalescing compact objects: black holes and neutron stars\\
(joint with DCOMP)\\
Yosef Zlochower Status of Black Hole-Black Hole  Simulations and Applications\\
Yuk Tung Liu Status of Neutron Star-Black Hole Simulations and Applications\\
Luciano Rezzolla Status of Neutron Star-Neutron Star Simulations and Applications\\

Women and Minorities in Multi-Messenger Astronomy of Gamma-Ray Bursts\\
(joint with DAP, COM and CSWP)\\
Laura Cadonati Gamma-Ray Burst Observations with LIGO\\
Enrico Ramirez-Ruiz Triggering Gamma-Ray Bursts\\
Ignacio Taboada Neutrino Messages from Gamma Ray Bursts\\

Panel Discussion: Women and Minorities in Gravity: Science and Career Paths\\
(joint with DAP, COM and CSWP)\\
Andrea Lommen Pulsar Timing and Gravitational Waves\\
Fotini Markopoulou Challenges in Quantum Gravity\\
Steve McGuire Materials Science in LIGO\\

Spanning the Gravitational Wave Spectrum\\
Vuk Mandic Gravitational wave astronomy using LIGO\\
Guido Mueller LISA - the other new window to the universe\\
Frederick Jenet Pulsar Timing to Search for nanoHz Gravitationa Waves\\

The Scientific Legacy of John Wheeler\\
(joint with FHP)\\
Kenneth Ford John Wheeler, 1933-1959: Particles and Weapons\\
Kip Thorne John Wheeler, 1952-1976: Black Holes and Geometrodynamics\\
Wojciech Zurek John Wheeler, 1976-1996: ``Law Without Law'' and Quantum Information\\

Precision Measurements in Gravity\\
Nergis Mavalvala Beyond the quantum limit in gravitational wave detectors\\
William Weber LISA Pathfinder: testing the limits of pure geodesic motion for gravitational wave observation in space\\
Michael Watkins The GRACE Mission\\

Developments in Quantum Gravity\\
James Hartle Einstein Prize Talk\\
Laurent Freidel TBD\\
Robert Wald TBD\\

In addition, there are a number of other invited sessions that will be of interest to some GGR members:

New Facilities in Particle Astrophysics I   (DAP)\\
Frank Krennrich AGIS-the Advanced Gamma-ray Imaging System\\
Joel Bregman Science Drivers for the International X-ray Observatory\\
Zeljko Ivezic LSST: the physics of the dark universe\\

New Facilities in Particle Astrophysics II  (DAP)\\
Stephan Meyer Probing Inflation with Cosmic Microwave Background Polarization\\
Jonathan Grindlay EXIST: Surveying Black Holes from the Early Universe to Local Galaxies\\
Thomas Prince LISA: Using Gravitational Waves to Observe the Universe\\

Early Science from the Fermi Gamma-Ray Space Telescope (DAP)\\
Charles A. Meegan Fermi Observations of Gamma-Ray Bursts\\
James Chiang Fermi/LAT Observations of the Extragalactic Gamma-Ray Sky\\
Troy A Porter Highlights of Galactic Scientific Results from the Large Area Telescope\\

New Eyes on the Universe I  (DAP)\\
Stefan Westerhoff New Results from the Pierre Auger Observatory\\
Martin Israel TIGER: Progress in Determining the Sources of Galactic Cosmic Rays\\
Todor Stanev Recent Progress and New Puzzles in Cosmic Ray Physics\\

New Eyes on the Universe II  (DAP)\\
Simon Swordy The VERITAS gamma-ray observatory: Recent observations and status\\
John Pretz Milagro's Survey of the TeV Gamma-Ray Sky\\
Teresa Montaruli Recent Results from IceCube\\

Dark Matter  (DPF)\\
Andrew Hime A DEAP \& CLEAN Program for the Direct Detection of Dark Matter\\
Mark Pearce Recent Results from PAMELA\\
Dan Hooper The Hunt for Dark Matter\\

Frontiers in Computational Astrophysics  (DCOMP)\\
Chris Fryer Stellar Core Collapse\\
Tiziana di Matteo The Interplay between Galaxy Evolution and Supermassive Black Hole Growth\\
Jonathan McKinney Accretion Flows around Compact Objects\\

Computational Astrophysics of Disks: From Black Holes to Planets  (DCOMP)\\
Charles Gammie Relativistic Magneto-Hydrodynamics and Black-Hole Accretion\\
Marina Romanova MHD Simulations of Disk-Star Interaction\\
Richard H. Durisen Radiative Hydrodynamics of Protoplanetary Disks and the Origin of Giant Planets\\

Nuclear Physics Connections with Astrophysics/Cosmology (DNP)\\
Sanjay Reddy Matter Under Extreme Conditions and Its Role in Explosive Astrophysical Phenomena\\
Paul Stankus Quark-Gluon Plasma: The Stuff of the Early Universe\\
Jorge Piekarewicz Neutron Stars and the PREX Experiment\\

Nuclear Physics in Astrophysics - From Stars to Stellar Explosions (DNP)\\
Dave Arnett Bethe Prize Talk: The Physics of Stars\\
D. Andrew Howell Dimming Metals: the Effect of Progenitor Metallicity on the 56Ni Yield and Luminosity of Type Ia Supernovae\\
Remco Zagers Test of Weak Reaction Rates of Importance for Late Stellar Evolution Using Charge-Exchange Reactions\\

Precision Measurements Impacting Cosmology (GPMFC)\\
Adam Reiss The Hubble Constant from the Hubble Space Telescope, Version 2.0\\
Maxim Pospelov Interacting dark energy in the lab: can we detect it?\\
Eric Adelberger Tests of Lorentz invariance using spinning electrons and the moon\\

\vfill\eject
\section*{\centerline
{we hear that \dots}}
\addtocontents{toc}{\protect\medskip}
\addcontentsline{toc}{subsubsection}{
\it we hear that \dots , by David Garfinkle}
\parskip=3pt
\begin{center}
David Garfinkle, Oakland University
\htmladdnormallink{garfinkl-at-oakland.edu}
{mailto:garfinkl@oakland.edu}
\end{center}

Jim Hartle is this year's winner of the APS {\it Einstein Prize} for 
Gravitational Physics.

Robert Caldwell, Steven Carlip, Chris Fryer, David Garfinkle, 
John Hughes, Vassiliki Kalogera, Frank Krennrich, 
Pablo Laguna, Eric Poisson, Norna Robertson, James Ryan 
and Michael Zucker
have been elected Fellows of the American Physical Society.

Hearty Congratulations!

\section*{\centerline
{400 years ago}}
\addtocontents{toc}{\protect\medskip}
\addcontentsline{toc}{subsubsection}{
\it 400 years ago, by David Garfinkle}
\parskip=3pt
\begin{center}
David Garfinkle, Oakland University
\htmladdnormallink{garfinkl-at-oakland.edu}
{mailto:garfinkl@oakland.edu}
\end{center} 

In a departure from our usual ``100 years ago'' format, we note that 
400 years ago Galileo first used a telescope for astronomy.  In celebration
of this monumental event 2009 has been designated the international year 
of astronomy.

\vfill\eject

\section*{\centerline
{The 24th Texas Symposium on Relativistic Astrophysics}}
\addtocontents{toc}{\protect\medskip}
\addtocontents{toc}{\bf Conference reports:}
\addcontentsline{toc}{subsubsection}{
\it The 24th Texas Symposium on Relativistic Astrophysics, 
by Scott Hughes}
\parskip=3pt
\begin{center}
Scott Hughes, Massachusetts Institute of Technology
\htmladdnormallink{sahughes-at-mit.edu}
{mailto:sahughes@mit.edu}
\end{center}

\bigskip

The Texas Symposium on Relativistic Astrophysics was held in
Vancouver, British Columbia, December 8th--12th of 2008.  This was
the 24th Texas meeting; the first was December, 1963 in Dallas, Texas.
The impetus for the first meeting was the discovery that certain odd
radio sources were cosmological, implying that they were extremely
luminous.  It quickly became clear that gravitational energy release
was a likely explanation for these sources (now called ``quasars''),
and so the meeting brought together researchers in relativity,
theoretical astrophysics and observational astronomy.  This is a
report on the plenary talks at Texas XXIV.

Virginia Trimble of UC Irvine reminded us of the meeting's history in
her talk ``A
Remarkable Achievement: Texas at 45.''  This was a comprehensive
overview of milestones reported at past Symposia (e.g., pulsars in
1968, Hawking radiation in 1972, the Hulse-Taylor pulsar in 1974,
supernova 1987a in 1988).  Virginia punctuated her talk by describing
problems getting speakers from various countries (e.g., Vitaly
Ginzburg was prevented from attending Texas I), issues with funding
(one year the conference was funded by Motorola, hoping for a tie-in with their
``Quasar'' televisions), and the slowly but (at least recently)
steadily increasing number of women speaking and organizing the
meeting.

The remainder of Day 1's plenary talks focused on recent observations
and facilities.  Shri Kulkarni of Caltech began by reviewing gamma-ray
bursts.  He first described successes of the recent past, such as
confirmation that most bursts are cosmological, and the discovery that
``long soft'' bursts emit in narrow jets.  Shri noted that there are
clouds on the horizon: Models explaining bursts are becoming somewhat
baroque, needing to invoke multiple jets and repeated shocks.
Finally, he described evidence that ``short hard'' bursts come from a
distinct population than the long soft bursts.  The emerging picture
indicates that short hard bursts are related to stellar remnants
rather than massive star collapse.

Peter Michelson of Stanford then discussed early results from the
Fermi Gamma-ray Telescope.  Formerly known as GLAST (the Gamma-ray
Large Area Space Telescope), it was renamed to salute Enrico Fermi's
contributions to astrophysics, and to acknowledge the role of
international collaboration in its development.  Fermi has already
gathered 100 times as many photons as EGRET, the previous all-sky
gamma-ray monitor.  One exciting result is the discovery of a pulsar
in the supernova remnant CTA1.  No remnant had been seen there before;
this appears to represent a pulsar which only emits in gamma rays.
Future surveys will look for dark matter annihilation, a prime mission
goal.

Day 1 concluded with Paul Sommers of Penn State discussing the Pierre
Auger cosmic ray observatory.  Pierre Auger consists of two sets sets
of Cerenkov detectors: A (completed) site in Argentina, with 1600
tanks spread over 3000 km$^2$, and a (planned) site in Colorado with
4400 tanks spread over 8000 km$^2$.  The southern site has important
science results, including confirmation of cosmic ray depletion above
the GZK (Greisen-Zatsepin-Kuzmin) cutoff, and an apparent correlation
with active galactic nuclei for rays with energy above $\sim 57 \times
10^{18}$ eV.  Unfortunately, the correlation appears strongest with
AGN that are not well covered by the southern hemisphere site.  Paul
was eager to develop the northern site to improve sky coverage.

Day 2 began with Manuela Campanelli of the Rochester Institute of
Technology describing ``Binary Black Hole Simulations,'' giving an
overview of the explosion of activity in numerical relativity in the
past few years.  Manuela focused on recent physics results, describing
the outstanding overlap between numerical waveforms and analytic
techniques, as well as describing what we have learned about mass and
spin evolution, and recoil imparted by gravitational-wave emission.
Sterl Phinney of Caltech then described ``Relativistic Astrophysics
with Gravitational Waves.''  Sterl reviewed how gravitational waves
can be used to formulate interesting strong-field gravity tests, probe
the nature and abundance of compact objects, and answer questions
in stellar evolution and stellar dynamics.  He concluded by noting
that many of these issues will be answered most clearly by
coordinating future gravitational-wave observations with telescope
observations in all bands, an appeal this audience was well positioned
to hear.

From compact objects we moved into cosmology.  Henk Hoekstra of Leiden
told us about ``Weak Lensing by Large-scale Structure,'' one of the
most promising techniques for weighing large groups of matter.  The
challenge here is to back out the ellipticity imposed by lensing shear
on galaxy images in the presence of a galaxy's intrinsic shape,
pixellation, blurring, and noise.  Statistics makes it possible: A
field of galaxies lensed by some structure will have common lensing
shear; other contributions should average out in a well-defined sense.
Wendy Freedman then discussed ``The Cosmic Distance Scale.''  Her talk
primarily described how recent measurements have improved our
knowledge of the distance to Cepheid variable stars in not-too-distant
galaxies.  Recall that Cepheids vary in luminosity with a fixed
period-luminosity relation.  By measuring the period of a distant
Cepheid, we learn its absolute magnitude.  From its measured apparent
magnitude, we then learn its distance.  Improving our knowledge of the
first ``rung'' on the cosmic distance ladder has a great impact on
{\it all} distances.  This work is largely responsible for pinning
down the Hubble constant to the roughly $10\%$ accuracy we use today.

Eiichiro Komatsu of the University of Texas opened Day 3 speaking on
``Non-gaussianity in the Cosmic Microwave Background.''  This was a
prize lecture, celebrating Eiichiro's win of the 2008 Young Scientist
Prize from the International Union of Pure and Applied Physics.  The
talk made it clear that this was a well-deserved prize; Eiichiro gave an
engaging, entertaining, and informative summary of how non-gaussianity
can arise and imprint itself on the CMB.  He demonstrated that the CMB
is Gaussian to better than $0.1\%$ --- in many ways a more stringent
constraint than our limits on spatial flatness.  Christof Pfrommer of
CITA then described ``High Energy Astrophysics in Galaxy Clusters.''
Chris's talk made it clear that clusters are extremely complicated
objects.  He made an excellent case that they will be interesting
targets for observing campaigns with Fermi; they should also be
excellent targets for low-frequency radio arrays, since cluster gas
will be shocked.  His talk introduced many of us to the German word
``gischt,'' the foam on top of a cresting wave, which he used to
describe emission from a shock front.

Reinhard Genzel of the Max-Planck-Institut for Extraterrestrial
Physics then brought us up to date on ``The Galactic Center.''  He
primarily focused on his group's work using the Very Large Telescope
in Cerro Paranal, Chile (with much credit also given to his main
competitor, Andrea Ghez's group at UCLA).  He now reports the mass of
the presumed black hole at the center of our galaxy as $3.95
\pm 0.06 \times 10^6\,M_\odot$, with most of the error due to uncertainty
in our knowledge of the distance to the galactic center (needed to
convert angular motions to physical motions).  This work requires
resolution of tens of milliarcseconds.  Reinhard described plans to
get to {\it micro}arcsecond resolution using optical interferometry at
the Very Large Telescope.  Day 3 concluded with a provocative talk by
Craig Hogan, now at Fermilab, entitled ``What's wrong with
cosmology?''  It appeared in the program as ``What's wrong with {\it
concordance} cosmology?'', but Craig decided to push things a little
further.  He cautioned the audience not to become too enamoured of
various ``isms,'' such as ``fundamentalism'' (only the deepest truths
matter!), ``giantism'' (if a 10 meter telescope is good, a 100 meter
telescope is better!), and ``futurism'' (10 years from now, data from
X will revolutionize this subject!).  It was a fun way to conclude
before our free afternoon.

Day 4 began with Cliff Will of Washington University telling us about
``Testing general relativity: A 30-year perspective and a view of the
future.''  The perspective part of the talk referred to the fact that
Texas IX occurred almost exactly 30 years earlier in Munich.  That
meeting featured the first presentation of the $\dot P$ measurement
from PSR 1913+16, with many concluding that testing general relativity
was done.  This viewpoint was quickly overthrown by work on fifth
forces (which rose and fell in the 1980s), string theory, extra
dimensions, Lorentz violation, dark matter and dark energy.  Cliff
noted that future tests are likely to be astronomical in nature,
pointing to work on imaging accretion disks around black holes, ever
more precise mapping of orbits, gravitational-wave astronomy, and
cosmology.  Ingrid Stairs of the University of British Columbia then
described ``The Double Pulsar,'' PSRJ0737-3039.  In introducing her,
Bill Unruh thanked whoever it was that prayed for this object, and
suggested that they keep at it.  Much of J0737's magic is due to the
serendipitous alignment of the binary to our line of sight: The
pulsars eclipse one another over their 2.4 hour orbit.  This allows us
to formulate tests that are far more stringent than those encountered
for almost any other astronomical objects.  For example, the Shapiro
delay due to pulses from one pulsar passing near the other agrees with
GR to within $0.05\%$.  The eclipsing geometry also lets us map the
magnetosphere of one of the neutron stars as pulses from the other are
affected by it.

We concluded this day with a pair of talks on cosmology.  Andrew
Liddle of the University of Sussex spoke on ``Inflation,'' reviewing
the subject and describing how well it is constrained at present.  A
particular point of interest is the index characterizing the spectrum
of scalar perturbations.  Inflation predicts $n_S$ slightly less than 1.
Andrew emphasized that, though the data currently indicates $n_S
\simeq 0.96$, one cannot really state with great confidence that $n_S
\ne 1$.  Tom Abel of Stanford then told us about ``Cosmic
Reionization.''  This talk was largely motivated by upcoming
low-frequency radio observations which will make 3-D tomographic maps
of hydrogen ionization in early structure formation.  Much like the
cosmic microwave background, these maps will trace the density and
distribution of matter, but do so over a range of redshifts rather
than at a single epoch.  Tom showed us that simulations of structure
growth can now trace matter evolution from a nearly isotropic initial
density field to produce protostars of a just a few Jupiter masses,
covering 23 orders of magnitude in density.

The meeting ended with a day of talks on cosmology.  Sean Carroll of
Caltech opened by describing ``Dark forces and dark energy.''  Noting
that this territory has been very well covered in various venues
recently, Sean quickly jumped to the punchline: Our universe appears
to be dominated by a cosmological constant, and the bulk of its matter
appears to be non-baryonic.  Sean described various mechanisms and
theories to explain these features, but noted that many alternates end
up introducing as many problems as they solve.  His advice was just to
do the observations and experiments and not pay too much attention to
theoretical objections.  Jonathan Feng of UC Irvine followed this by
discussing ``Dark matter candidates and signals.''  Jonathan
emphasized, much as Sean did, that there is an enormous range of
possibilities, with very few constraints.  He described the various
experiments that aim to directly detect dark matter particles, and
gave a brief summary of recent observations (DAMA's annual
modulations; the positron excess seen by the PAMELA and ATIC
satellites) which {\it may} be related to dark matter.  Unfortunately,
it is hard to find a model which explains {\it all} of these data!
Phil Hopkins of UC Berkeley then told us about ``Quasars, Feedback,
and Galaxy Formation.''  This was a true gastrophysics talk, examining
the ways in which quasars are fueled, how their output acts back on
their host galaxy (``feedback''), and what this implies for the
evolution of galaxy structure and quasar luminosity function with
cosmic time.

Lawrence Krauss, now at the University of Arizona, wrapped up the
meeting with ``The Future of the Universe and the Future of
Cosmology,'' or ``Our Miserable Future.''  He was quite pessimistic
about ever understanding dark energy.  Though detection of only a tiny
deviation from a pure $\Lambda$ universe would immediately falsify the
cosmological constant hypothesis, we are unlikely to be able to
measure a tiny deviation.  He made the interesting observation that if
the cosmological constant hypothesis is correct, then we are evolving
toward deSitter's static universe: Everything beyond a relatively
nearby horizon will move out of causal contact.  Future cosmologists
(who he described as ``lonely and ignorant ... but dominant'') will
have no need to develop ``cosmology'' as we now know it, since the
whole universe will just be our galaxy.  These cosmologists will be
led, by the best data available to them, to a totally incorrect
fundamental model of the universe --- a point he suggests we bear in
mind when examining our data today.

In her opening remarks, Virginia Trimble noted that when the first
Symposium was proposed, Engelbert Schucking asked his co-organizers,
``What the hell is {\it relativistic astrophysics}?''  Judging by the
talks presented here in Vancouver, it appears to be astrophysics in
which general relativity plays a crucial role.  It's worth recalling
the concluding remarks of Tommy Gold at the first Texas meeting:
``... here we have a case ... that the relativists with their
sophisticated work were not only magnificent cultural ornaments but
might actually be useful to science! ... It is all very pleasing, so
let us hope that it is right.  What a shame it would be if we had to
go and dismiss all the relativists again.''  Texas XXIV shows that
there's not much danger of such dismissal anytime soon.  I'm looking
forward to Texas XXV.

\vfill\eject

\section*{\centerline
{Loop Quantum Cosmology Workshop}}
\addtocontents{toc}{\protect\medskip}
\addcontentsline{toc}{subsubsection}{
\it Loop Quantum Cosmology Workshop, by Parampreet Singh}
\parskip=3pt
\begin{center}
Parampreet Singh, Perimeter Institute for Theoretical Physics
\htmladdnormallink{psingh-at-perimeterinstitute.ca}
{mailto:psingh@perimeterinstitute.ca}
\end{center}

Loop quantum cosmology (LQC) is a non-perturbative quantization of
symmetry reduced spacetimes, based on loop quantum gravity. In
recent years it has emerged as a promising arena for quantum
gravity of simple yet rich models, offering interesting physical
implications. These include insights on the resolution of
spacelike singularities and ramifications of quantum gravity on
the physics of the very early universe. The LQC workshop at
the Institute of Gravitation and the Cosmos at Penn State was the
first such meeting in this field. It was a 3-day event, October
23-25, 2008, bringing more than 40 participants including many LQC
experts  and young researchers together under the same roof. A
highlight of this workshop was its attendance and talks by experts
from other fields who provided insights from an inter-disciplinary
perspective. The meeting took place in an academically stimulating
atmosphere with a large number of open discussions during the
talks and in special
discussion sessions. \\

There were a total of 21 talks including three reviews on LQC;
some were of a longer duration (45 minutes to 1 hour) and others,
shorter seminars (30 minutes). The workshop was divided into
different themes for each day. On the first day it focused on the
nature of spacetime near classical singularities and homogeneous
models in LQC. Day 2 dealt with quantization of inhomogeneous
models, effects of quantum geometry on quantum fields and
effective equations. Day 3 focused on physical implications and
phenomenological models. An important part of the schedule were
the daily focused discussion sessions (90 minutes each). These
sessions were also used for
very short voluntary talks by the participants.  \\

The workshop started with opening remarks by Abhay Ashtekar (Penn
State) who welcomed all participants and gave a brief introduction
about the progress in the field. He stressed the importance of
distinguishing between results which have been rigorously proved
and established from those which are preliminary in nature. The
morning session on the first day was on {\it Classical Singularities}
in which Beverly Berger (National Science Foundation) gave a
review of various mathematical and numerical results on the approach
to singularities in classical general relativity, obtained by
different groups in recent years. She pointed out that there is
ample emerging evidence in support of the BKL conjecture. Adam
Henderson (Penn State) then reported on a formulation of the BKL
conjecture that is well-adapted to LQC. Frans Pretorius
(Princeton) discussed recent results on Ekpyrotic/Cyclic models
which show that the ultra stiff equation of state for the radion
field suppresses the chaotic mixmaster behavior near the big bang.
These results indicate that spacetime near the big bang
singularity may be homogeneous. The afternoon session focused on
{\it Homogeneous Loop Quantum Cosmology} in which Abhay Ashtekar
gave a review, summarizing various  recent developments on 
singularity resolution. He also gave an update on new results
which include a quantum bounce at the Planck scale in inflationary
models. In this session, Edward Wilson-Ewing (Penn State)
presented the first consistent loop quantization of Bianchi-I
spacetime and discussed emerging exciting physics which shares the
feature of quantum bounce with isotropic models. In the discussion
session, Frans Pretorius stressed exchange of ideas between LQC
and Ekpyrotic models to achieve singularity resolution in the latter.\\

On the second day, the morning session focused on {\it
Inhomogeneous Loop Quantum Cosmology}. Jerzy Lewandowski (Warsaw
University) gave a review of results on the effects of quantum
geometry on quantum fields. Alejandro Corichi (UNAM, Morelia) and
Guillermo A. Mena Marugan (IEM (CSIC), Madrid) gave short reviews
of the Fock and Hybrid quantizations of the Gowdy models
respectively. The afternoon session talks were on the {\it
Effective Equations} in which Martin Bojowald (Penn State) gave an
overview of one of the approaches to obtain effective dynamics.
The discussion session on the second day featured two short talks
on very interesting results. The first one by Victor Taveras (Penn
State) was on modifications to Friedman dynamics in LQC. The
second talk was by Tomasz Pawlowski (IEM (CSIC), Madrid) who
presented results using effective dynamics in the Hybrid
quantization of Gowdy models. He discussed numerical simulations
performed by the Madrid group which show the presence
of a quantum bounce in the presence of inhomogeneities.\\

The final day of the workshop  focused on {\it Loop quantum
gravity phenomenology, Inflation and CMB}.  Parampreet Singh
(Perimeter) gave a review of recent results obtained by different
groups and discussed fate of singularities using effective
dynamics in LQC. Mairi Sakellariadou (Kings College) gave an
overview of the numerical results obtained from lattice refinement
techniques. This was followed by a talk by William Nelson (Kings
College)  on phenomenological aspects of these techniques in dark
energy models. The afternoon session featured short talks on
preliminary results on issues inspired by loop quantum gravity.
These included effects on cosmological perturbations (Michal
Artymowski (Warsaw University) and Gianluca Calcagni (Penn State)) and
implications of treating the Barbero-Immirzi parameter as a field
(Nico Yunes (Princeton)). In this session Tirthabir Biswas (Penn
State) presented ideas to obtain a scale invariant spectrum of
perturbations  using the string gas mechanism. In the concluding
session, participants discussed various new results presented
during the workshop, assessed what has been achieved so
far and open directions which need to pursued in the near future.\\

The workshop dinner took place in the Nittany Lion Inn, one of the
historic hotels of the United States. Participants used this warm
occasion to discuss the possibilities to organize another workshop
focused on LQC  in near future. External experts commented that
the workshop was a wonderful opportunity to get updates and
clarifications on recent developments. Because of the emphasis on
discussions, younger participants found it
especially rewarding. Online proceedings are available at\\
\htmladdnormallink
{\protect {\tt {http://www.gravity.psu.edu/events/loop\_quantum\_cosmology/proceedings\_lqc.shtml
}}}
{http://www.gravity.psu.edu/events/loop_quantum_cosmology/proceedings_lqc.shtml}

\end{document}